\documentstyle[12pt,epsfig,prc,preprint,aps]{revtex}
\tightenlines

\begin{document}
\preprint{\scriptsize{To appear in Phys.\ Rev.\ C {\bf 56} (Nov. 1997)}}

\draft

\title{The role of retardation in 3-D relativistic equations}

\author{A. D. Lahiff and I. R. Afnan}

\address{Department of Physics,
        The Flinders University of South Australia,
        GPO Box 2100, Adelaide 5001, Australia}

\date{\today}

\maketitle

\begin{abstract}

Equal-time Green's function is used to derive a three-dimensional integral
equation from the Bethe-Salpeter equation. The resultant equation, in the 
absence of anti-particles, is identical to the use of time-ordered diagrams, 
and has been used within the framework of $\phi^2\sigma$ coupling to study
the role of energy dependence and non-locality when the two-body potential
is the sum of $\sigma$-exchange and crossed $\sigma$ exchange. The
results show that non-locality and energy dependence make a substantial
contribution to both the on-shell and off-shell amplitudes. 

\end{abstract}


\section{Introduction}\label{sec.1}

With the advent of Quantum Chromodynamics (QCD) as the fundamental 
underlying theory of strong interaction, it would be nice to relate the
Lagrangian used in the nucleon-nucleon interactions with a 
meson-baryon effective Lagrangian extracted from QCD. 
Although it is not possible at present to write such an effective 
Lagrangian with all the coupling of the meson to the baryon 
predetermined  by QCD, we can start with a Lagrangian that preserves 
the symmetries of  QCD, {\it e.g.} chiral symmetry. At this stage the coupling
constants will have physical significance in that they can be related to 
QCD parameters. On the other hand, a determination of the coupling 
constants in an effective chiral Lagrangian\cite{ORK96} from the 
experimental data, {\it e.g.} the nucleon-nucleon phase shifts, could 
be used to test models of QCD.

Modern nucleon-nucleon interactions\cite{MHE87,SKT94} based on meson 
exchange have achieved the remarkable success of fitting the 
`experimental' phase shifts with a $\chi^2$ per data of approximately 
one\cite{SKT94}. 
These interactions invariably start with a Lagrangian that includes the 
coupling of the nucleon to a set of 
mesons, with the coupling constants of the mesons to the nucleon as 
parameters to be adjusted to fit the experimental data. To determine the 
phase shifts for a given Lagrangian, we need to: 
(i)~Define the two-body equation to be solved for the scattering amplitude. 
(ii)~Define the kernel or potential for this two-body equation. 
In an ideal world, to maintain covariance, we need to solve the 
Bethe-Salpeter (BS) equation\cite{SB51} with the kernel
being the sum of all two-particle irreducible Feynman diagrams for 
nucleon-nucleon scattering. In practice this is not simple, and 
the standard procedure has been: 
(i)~To replace the four-dimensional Bethe-Salpeter equation by a 
three-dimensional (3-D) equation that reduces to the 
Lippmann-Schwinger equation\cite{LS50} in the non-relativistic limit. 
(ii)~Approximate the potential for this 3-D equation by the sum of all single 
meson exchanges or one- plus two-meson exchanges. 
(iii)~To use the resultant potential in three or more nucleon systems 
({\it i.e.} the Schr\"odinger equation), reduce 
the momentum and energy dependence of the potential in order to generate 
a local coordinate space potential.

The aims of the present investigation are to test, in a scalar theory, the 
effect on the coupling constant of: 
(i)~Replacing the Bethe-Salpeter equation by a corresponding 3-D equation. 
(ii)~The subtraction of the anti-particles' contributions from the kernel.
(ii)~The removal of the non-locality and energy dependence in the potential 
to allow for a coordinate space local potential. 
This will give us a qualitative measure of the error in the 
coupling constant resulting from the standard approximation used in 
generating local nucleon-nucleon potentials.

In principle there are an infinity of relativistic 3-D equations that satisfy 
the same unitarity conditions as the Bethe-Salpeter equations\cite{KL74}. 
However, in practice there are four equations most commonly used in 
nucleon-nucleon scattering. Three result as a direct reduction of the
Bethe-Salpeter equation. These are: 
(i)~The Blankenbecler-Sugar \cite{BbS66} equation in which the 
relative energy is set to zero. 
(ii)~The Gross \cite{G82} equation in which  one of the particles 
is on-mass-shell. 
(iii)~The Klein\cite{K53} equation in which the relative energy is 
integrated out. This latter equation has been used in recent years in 
conjunction with time ordered perturbation theory to determine the 
two-meson exchange nucleon-nucleon potential. The difference between these
three equations is the treatment of the off-mass-shell degree of freedom
present in the BS equation.
The fourth equation is based on Hamiltonian dynamics\cite{KP91} with the
connection to the field theoretic Lagrangian made via the Okubo projection
method\cite{Ok54} as implemented by Fuda\cite{Fu92}. In this equation the
nucleons are on-mass-shell. 

In Sec.~II we will derive the Klein equation from the equal-time Green's 
function\cite{LT63}, and in this way establish a systematic
procedure for improvement on the standard potential in the Klein equation 
as suggested by Phillips and Wallace\cite{PW96}. Although we have reduced
the dimensionality of the equation from 4-D to 3-D, the kernel still
includes the contribution from anti-particles. 
In Sec.~III we examine the kernel of the Klein equation.  Here we 
find that the contribution of anti-particles, at the level of one meson
exchange, can be significant for sufficiently large coupling constants.
At a coupling constant that gives a binding energy comparable to that
of the deuteron, the contribution from anti-particles can be neglected.
The neglect of the anti-particle contribution reduces the kernel 
of the Klein equation to that resulting from time-ordered perturbation 
theory as has been implemented for the nucleon-nucleon system in both 
the Bonn and Nijmegen potentials.
We then proceed to define the potential arising from the 
exchange of two mesons in the no-anti-particle (NAP) approximation.
In the Klein approach, with NAP, the one- plus two-meson 
exchange potential is both non-local and energy dependent, and to derive 
the corresponding local potential we need to remove these dependencies. 
In Sec.~IV we compare the results of the BS, Klein-NAP and the energy 
independent potentials with different levels of non-locality. We find
that although the BS and Klein-NAP are in reasonable agreement, the
removal of non-locality and energy dependence could be too severe. 
The effect of these approximations and the introduction of a form 
factor on the value of the coupling constant 
and the off-shell behavior of the potential is examined. 
Finally, in Sec.~V we discuss some of the possible implications 
of our results for the nucleon-nucleon interaction.

\section{Theory}\label{sec.2}

To establish the relation between the BS and Klein equations, we derive the 
Klein equation\cite{K53} from the equal-time Green's function\cite{LT63}. 
In this way, we establish the relation between the four 
dimensional Bethe-Salpeter equation\cite{SB51} and the three dimensional 
Klein equation in the form recently proposed by Phillips and Wallace\cite{PW96}
which allows for a systematic improvement in the kernel of the Klein equation
to reproduce the result of the BS equation. 

The two-body Green's function or four-point  function is given  
in terms of the vacuum expectation value of the time ordered 
product of  fields as\cite{IZ85}:
\begin{equation}
G(x,y;x',y') \equiv \langle 0|\,T(\psi(x)\,\psi(y)\,
\bar{\psi}(x'),\bar{\psi}(y')\,|0\rangle               \label{eq:1}
\end{equation}
The equal-time Green's function is then defined as\cite{LT63}:
\begin{equation}
G(x_0{\bf x},x_0{\bf y};x'_0{\bf x}',x'_0{\bf y}') \equiv
\int \,dy_0\,dy'_0\ \delta(x_0-y_0)\,\delta(x'_0-y'_0)
\ G(x,y;x',y')                                          \label{eq:2}
\end{equation}
where we have written the four-vectors using the notation $x=(x_0,{\bf x})$. To 
simplify matters, we need to write the coordinate space Green's function in
terms of the momentum space Green's function. This can be achieved by first
introducing the relative and center of mass momenta as:
\widetext
\begin{equation}
k=\frac{1}{2}(p_1-p_2) \quad\mbox{and}\quad P = p_1+p_2    \label{eq:3}
\end{equation} 
\narrowtext
then $ d^4p_1\,d^4p_2 = d^4P\,d^4k$. The equal-time Green's function can be 
now be written as
\begin{eqnarray}
G(x_0{\bf x},x_0{\bf y};x'_0{\bf x}',x'_0{\bf y}') & = & 
\int\,d^4P\,d^3k\,d^3k'\,e^{iP_0(x_0-x'_0)}\,e^{-i{\bf P}
\cdot[({\bf x}+{\bf y})/2 -
({\bf x}' + {\bf y}')/2]} \nonumber \\
& &\qquad\times e^{-i{\bf k}\cdot({\bf x}-{\bf y})}\,
\tilde{G}({\bf k},{\bf k}';P)\,
e^{i{\bf k}'\cdot({\bf x}'-{\bf y}')}                   \label{eq:4}
\end{eqnarray}
where $\tilde{G}({\bf k},{\bf k}';P)$, the `equal-time' Green's function in
momentum space, is given by
\begin{equation}
\tilde{G}({\bf k},{\bf k}';P) = \int\limits^{+\infty}_{-\infty}
\,dk_0\,dk'_0\ 
G(k_0{\bf k},k'_0{\bf k}';P) 
\equiv \langle\,G\,\rangle\ .                         \label{eq:5}
\end{equation}
We note that this `equal-time' Green's function is {\it not} a function of the 
`relative energy' $k_0$, and has the potential of setting the framework for a
three-dimensional  integral equation for both the Green's function and the
scattering amplitude or $T$-matrix.

The two-body Green's function defined in Eq.~(\ref{eq:1}) satisfies the 
BS equation
\begin{eqnarray}
G &=& G_0 + G_0\,K\,G \nonumber \\
  &=& G_0 + G_0\,T\,G_0                                \label{eq:6}
\end{eqnarray}
where $G_0$ is the free two-particle Green's function, the product of the
Feynman propagators for the two particles, and $T$ is the $T$-matrix for 
two-particle scattering. In Eq.~(\ref{eq:6}), $K$ is the potential in 
the BS equation and consists of the sum of all two-particle irreducible
Feynman diagrams that contribute to the Green's function. 
Using the second line in Eq.~(\ref{eq:6}) to iterate the first line, 
we can to write the BS equation for the $T$-matrix as
\begin{equation}
T = K + K\,G_0\,T                              \ .      \label{eq:7}
\end{equation}
The aim of this section is to write an approximation to 
Eq.~(\ref{eq:7}) that
is an integral equation in three-dimensions. There are an infinity of such
equations\cite{KL74}, and to find the optimum one is not the aim of the 
present study. 

The unique feature of the Klein equation is that the 
potential is assumed to be independent on the `relative' 
energy $k_0$, and as a result the `relative' energy integration,
in the integral equation, can be carried through resulting in a 3-D 
equation.
This suggests that one may develop an approximation scheme based on the 
idea that the potential $K$ can be divided in two parts\cite{PW96}. 
The first part $K_1$ is independent of the `relative' energy, i.e.,
\begin{equation}
K_1(k,k';P) = K_1({\bf k},{\bf k'};P)\ ,       \label{eq:8}
\end{equation}
and the rest, $K_2=K - K_1$, has all the `relative' energy dependence. 
The problem now reduces to finding an optimum $K_1$ such that the solution
to the 3-D integral equation is a good approximation to the solution of the
BS equation, and to define a systematic procedure for improving the results of
the Klein equation.

Let us first assume that $K_2=0$, then the `equal-time' Green's function
can be written as
\begin{eqnarray}
\langle\,G\,\rangle &=& \langle\,G_0\,\rangle 
                   + \langle\,G_0\,T\,G_0\,\rangle    \nonumber\\
              &=& \langle\,G_0\,\rangle 
                   + \langle\,G_0\,K_1\,G\,\rangle\ ,\label{eq:9}
\end{eqnarray}
where
\begin{equation}
\langle\,A\,\rangle \equiv \tilde{A}({\bf k},{\bf k'};P) \equiv
\int\limits^{+\infty}_{-\infty}dk_0\,dk'_0\,A(k,k';P)\ .\label{eq:10}
\end{equation}
However, since $K_1$ does not depend on the `relative' energy, it is simple
to show that $\langle\,G_0\,K_1\,G\,\rangle = \langle\,G_0\,\rangle\,K_1
\,\langle\,G\,\rangle$ and as a result we can write
\begin{eqnarray}
\langle\,G_0\,T\,G_0\,\rangle &=& \langle\,G_0\,\rangle
                                  \,K_1\,\langle\,G\,\rangle  \nonumber\\
   &=& \langle\,G_0\,\rangle\,K_1\,\left\{ \langle\,G_0\,\rangle +
     \langle\,G_0\,T\,G_0\rangle\right\}   \ .         \label{eq:11}
\end{eqnarray}
This allows us to write a 3-D integral equation, the Klein equation, 
in which $K_1$ is the potential, as
\begin{equation}
T_1 = K_1 + K_1\,\langle\,G_0\,\rangle\,T_1   \ ,     \label{eq:12}
\end{equation}
where the amplitude $T_1$ is the `equal-time' $T$-matrix, and
is defined in terms of the BS $T$-matrix as
\begin{equation}
T_1 \equiv \langle\,G_0\,\rangle^{-1}\ \langle\,G_0\,T\,G_0\,\rangle\ 
\langle\,G_0\,\rangle^{-1}       \ .                  \label{eq:13}
\end{equation}
At this stage all we have established is that if $K_1$ is independent of the
relative energy, then we have the 3-D equation first proposed by Klein\cite{K53}.

The potential in the BS equation is the sum of all two-particle irreducible 
diagrams that contribute to the amplitude, and in general this potential 
depends on the `relative' energy. In this case $K_2$ is not zero, and 
we need to define an optimum $K_1$. The Green's function for the potential
$K_2$ is given by
\begin{equation}
G_{K_2} = G_0 + G_0\,K_2\,G_{K_2}
        = G_0 + G_0\,\left(K - K_1\right)\,G_{K_2} \ ,\label{eq:14}
\end{equation}
and the corresponding `equal-time' Green's function is given by
\begin{equation}
\langle\, G_{K_{2}}\,\rangle = \langle\, G_{0}\,\rangle + 
\langle\,G_{0}\,K\, G_{K_{2}}\,\rangle - 
\langle\,G_{0}\,\rangle\,K_{1}\,\langle\, G_{K_{2}}\,\rangle 
\ .                                                   \label{eq:15}
\end{equation}
If we now define $K_{1}$, the potential that is independent of the 
`relative energy', in terms of the kernel of the BS equation as
\begin{equation}
K_{1} \equiv \langle\, G_{0}\,\rangle^{-1}\, 
        \langle\,G_{0}\,K\, G_{K_{2}}\,\rangle 
        \,\langle\, G_{0}\,\rangle^{-1}  \ ,          \label{eq:16}
\end{equation}
then Eq.~(\ref{eq:15}) can be solved for the `equal-time' Green's 
function for the potential $K_{2}$ with the simple solution that
$\langle\,G_{K_2}\,\rangle = \langle\,G_0\,\rangle$ i.e. the 
`equal-time' Green's function for the potential $K_2$ is identical 
to the `equal-time' free Green's function, provided $K_{1}$ is defined 
as in Eq.~(\ref{eq:16}).

The Green's function for the full BS kernel $K$ can now be written in 
terms of $K_{2}$ as
\begin{eqnarray}
G &=& G_0 + G_0\,T\,G_0 \nonumber \\
  &=& G_{K_2} + G_{K_2}\,K_1\,G     \ ,               \label{eq:17}
\end{eqnarray} 
with the corresponding `equal-time' Green's function given by
\begin{eqnarray}
\langle\,G\,\rangle &=& \langle\,G_0\,\rangle + \langle\,G_0\,T\,G_0\,
                                                   \rangle \nonumber \\
  &=& \langle\,G_{K_2}\,\rangle 
  + \langle\,G_{K_2}\,K_1\,G\,\rangle \ .             \label{eq:18}
\end{eqnarray} 
We are now in a position to write the `equal-time' amplitude as
\begin{eqnarray}
\langle\,G_0\,T\,G_0\,\rangle &=& \langle\,G_{K_2}\,K_1\,G\,\rangle
= \langle\,G_{K_2}\,\rangle\,K_1\,\langle\,G\,\rangle  \nonumber \\
&=& \langle\,G_0\,\rangle\,K_1\,\left\{\,\langle\,G_0\,\rangle + 
\langle\,G_0\,T\,G_0\rangle\,\right\}\ .              \label{eq:19}
\end{eqnarray}
This result again gives us the Klein equation as defined in 
Eq.~(\ref{eq:12}) with the potential in this case given by 
Eq.~(\ref{eq:16}). 
Clearly, we cannot determine this potential for the full BS kernel with
$G_{K_2}$ a solution of Eq.~(\ref{eq:14}). This would be equivalent to
solving the full BS equation with no approximations to the kernel. 
As a first approximation, we could expand the kernel of the BS equation 
in powers of the coupling constant, and at the same time iterate 
Eq.~(\ref{eq:14}) to keep all terms of the same order as in the BS kernel.
This will effectively give us an expansion of the potential in the Klein
equation as a power series in the coupling constant. 
We will see in the next section, for Yukawa type coupling, that to 
order $g^2$ this potential corresponds to single meson exchange, 
while to order $g^4$, $K_1$ will include one- and two-meson exchanges. 
The question then is: how do the results of this three-dimensional 
equation (the Klein equation) compare with the results from the 
BS equation when the potential is calculated to the same order?

The definition of the `equal-time' potential in Eq.~(\ref{eq:16}) allows
for a systematic way of calculating the amplitude within the framework
of a three-dimensional equation. In addition, this equation includes 
both the positive and negative energy component of the Feynman propagators.
This result is identical to that used by Phillips and Wallace for 
the two-body bound state problem\cite{PW96}. One can show that in the 
absence of negative energy states, this expression for the potential is 
equivalent to the result of time-ordered perturbation theory\cite{KB94,P95}.
Since the $S$-matrix resulting Eq.~(\ref{eq:12}) with $K_1$ defined in 
Eq.~(\ref{eq:16}) is identical to the $S$-matrix from the original BS equation,
we have maintained covariance, but it is not manifest covariance. Furthermore, 
since truncation in both $K$ and $G_{K_2}$ are carried through at the field
theory level, we expect covariance to be maintained. 

Finally, to fourth order in the coupling constant, and in the absence of 
anti-particles, this potential is identical to that used to include 
two pion exchange in the nucleon-nucleon interaction before any 
approximation is required to transform the potential to coordinate 
space\cite{R91,RS96}. This therefore can form the basis of estimating
the approximate magnitude of the errors made in going from the BS equation
to a coordinate space local potential  used in the Schr\"odinger equation 
for a given Lagrangian.

\section{The two-body potential}\label{sec.3}

To examine the approximations needed to reduce the kernel of the
BS equation to a local potential in coordinate space,
we need to define a Lagrangian. To avoid the problems with spin and 
isospin, especially at the BS level with crossed meson exchange, 
we have considered a $\phi^2\sigma$ theory, {\it i.e.} 
a Lagrangian of the form
\begin{equation}
{\cal L} = \frac{1}{2}\,\left[\,\partial^\mu\phi\,\partial_\mu\phi 
           - m^2\phi^2\,\right] 
       + \frac{1}{2}\,\left[\,\partial^\mu\sigma\,\partial_\mu\sigma 
           - \mu^2\sigma^2\,\right]
         - g\,\phi^2\sigma\ .                       \label{eq:20}
\end{equation}
In the nucleon-nucleon interaction the kernel of the BS equation is
truncated to include the one pion exchange and the crossed two pion
exchange. To compensate for this truncation one includes the exchange
of all mesons with a mass less than 1~GeV. The motivation is that the
heavy meson exchange will model all higher order diagrams that have
been excluded~\cite{SR97}. With this approximation in mind, we make use 
of the above Lagrangian to define the kernel of the BS equation in 
terms of $\sigma$-exchange and the crossed $\sigma$-exchange. 
In addition, we will define the potential in the Klein equation 
to be an approximation to $K_1$, and the order of this approximation to be 
determined by our approximation to the kernel of the BS equation,  
{\it i.e.} the potential in both equations is taken to the same order
in the coupling constant. In this way we  expect the mechanism included 
in both the BS and Klein equations to be the same, and the difference is
the result of using a 3-D equation in place of the 4-D equation.

With the above scheme in mind, we first consider the potential in the 
Klein equation to order $g^2$. In this case we include the kernel of BS
equation to order $g^2$ and take $G_{K_2}\approx G_0$. This gives the 
single $\sigma$-exchange potential as:
\begin{equation}
K^{(2)}_1 = \langle\,G_0\,\rangle^{-1}\ \langle\,G_0\,I^{(2)}\,G_0\,
\rangle\ \langle\,G_0\,\rangle^{-1}   \ ,           \label{eq:21}
\end{equation}
where $I^{(2)}$, the $\sigma$-exchange amplitude as employed in the BS 
equation, is given by
\begin{equation}
I^{(2)}(k,k') = \frac{g^2}{(k-k')^2-\mu^2+i\epsilon} 
\ ,                                                  \label{eq:22}
\end{equation}
with the relative four-momentum $k=(k_0,{\bf k})$.
To evaluate the `equal-time' matrix elements $\langle\,G_0\,\rangle$ and 
$\langle\,G_0\,I^{(2)}\,G_0\,\rangle$, we need to perform the relative energy
integration in both the initial and final states (see Eq.~(\ref{eq:10})). 
For this we decompose the product of the Feynman propagators for the two 
$\phi$ fields, in the center of mass, in terms of their positive and
negative energy components, {\it i.e.} 
\begin{eqnarray}
G_0(k,P)&=& \frac{i}{(2\pi)^4}\ 
           \frac{1}{(\frac{\sqrt{s}}{2}+k_0)^2-E_k^2+i\epsilon}\ 
           \frac{1}{(\frac{\sqrt{s}}{2}-k_0)^2-E_k^2+i\epsilon}\ ,
                                                  \nonumber\\
   &=&G_0^{++} + G_0^{+-} + G_0^{-+} + G_0^{--}\ . \label{eq:23}
\end{eqnarray}
By performing the relative energy integration, we reduce the `equal-time' 
free Green's function to:
\begin{eqnarray}
\langle\,G_0\,\rangle &=& \frac{1}{(2\pi)^3}\ \frac{1}{E_k}\ 
                   \frac{1}{s - (2E_k)^2 + i\epsilon} \nonumber\\
  &=& \langle\,G_0^{++}\,\rangle + \langle\,G_0^{--}\,\rangle 
  \ ,                                              \label{eq:24}
\end{eqnarray}
where
\begin{eqnarray}
\langle\,G_0^{++}\,\rangle &=& \frac{1}{(2\pi)^3}\,\frac{1}{(2E_k)^2}\,
           \frac{1}{\sqrt{s}-2E_k + i\epsilon} \nonumber \\
\langle\,G_0^{--}\,\rangle &=& - \frac{1}{(2\pi)^3}\,\frac{1}{(2E_k)^2}\,
           \frac{1}{\sqrt{s}+2E_k - i\epsilon}\ .\label{eq:25}
\end{eqnarray}
We note here that the free `equal-time' Green's function is a function 
$s=P^2$ and not $\sqrt{s}$ as in a number of three-dimensional equations. It 
is only when we divide this Green's function into its positive and
negative energy components, and neglect the negative energy component, 
do we get the dependence on $\sqrt{s}$.

To perform the relative energy integration on the Green's function for
$\sigma$-exchange, ({\it i.e.} $G_0\,I^{(2)}\,G_0$) we again take advantage 
of the decomposition of the Green's function 
in terms of a positive and negative energy component to write
\begin{eqnarray}
\langle\,G_0\,I^{(2)}\,G_0\,\rangle &=& \int\limits^{+\infty}_{-\infty}\  
       dk_0\,dk'_0\ G_0(k,P)\,I^{(2)}(k,k')\,G_0(k',P)\nonumber \\
    &=& \sum_{\alpha\beta}\,\sum_{\gamma\delta}\ 
    I^{(2)}_{\alpha\beta;\gamma\delta}({\bf k},{\bf k'};P)   
\,                                                \label{eq:26}
\end{eqnarray}
where 
\begin{equation}
I^{(2)}_{\alpha\beta;\gamma\delta}({\bf k},{\bf k'};P) 
= \langle\,G_0^{\alpha\beta}\,I^{(2)}\,G_0^{\gamma\delta}\,\rangle
          \ ,                                     \label{eq:27}
\end{equation}
with $\alpha,\beta,\gamma,\delta=+,-$. Making use of the symmetry of
the integrals $I^{(2)}_{\alpha\beta;\gamma\delta}({\bf k},{\bf k'};P)$
under the exchange of indices and momenta to reduce the number of integrals, 
we can write the `equal-time' Green's function for $\sigma$-exchange as:
\begin{eqnarray}
\langle\,G_0\,I^{(2)}\,G_0\,\rangle 
    &=& \left[\,\langle\,G_0^{++}\,\rangle 
      + \Delta\,\right]\,d_+\,\left[\,\langle\,G_0^{++}\,\rangle 
      + \Delta\,\right]            \nonumber \\
    &&\ +\ \left[\,\langle\,G_0^{++}\,\rangle 
      + \Delta\,\right]\,d_+\,\left[\,\langle\,G_0^{++}\,\rangle 
      + \Delta\,\right]             \nonumber \\
    &&\ +\ \left[\,\langle\,G_0^{--}\,\rangle 
      + \Delta\,\right]\,d_-\,\left[\,\langle\,G_0^{--}\,\rangle 
      + \Delta\,\right]             \nonumber \\
    &&\ +\ \left[\,\langle\,G_0^{--}\,\rangle 
      + \Delta\,\right]\,d_-\,\left[\,\langle\,G_0^{--}\,\rangle 
      + \Delta\,\right]             \nonumber \\
    &&\ -\ 2 \langle\,G_0^{++}\,\rangle\,d_0\,\langle\,G_0^{--}\,
      \rangle - 2 \langle\,G_0^{--}\,\rangle\,d_0\,\langle\,G_0^{++}\,
      \rangle  \ ,                               \label{eq:28}
\end{eqnarray}
where
\begin{eqnarray}
d_+({\bf k},{\bf k'}) &=& \frac{g^2}{2\omega}\,
       \frac{1}{\sqrt{s}-E_k-E_{k'}-\omega+i\epsilon}  \nonumber \\
d_-({\bf k},{\bf k'}) &=& - \frac{g^2}{2\omega}\,
       \frac{1}{\sqrt{s}+E_k+E_{k'}+\omega-i\epsilon}   \nonumber \\
d_0({\bf k},{\bf k'}) &=& \frac{g^2}{2\omega}\,
       \frac{1}{E_k+E_{k'}+\omega}        \ ,       \label{eq:29}
\end{eqnarray}
and
\begin{equation}
\Delta({\bf k},{\bf k'}) 
= -\frac{1}{(2\pi)^3}\,\frac{1}{(2E_k)^2}\,\frac{1}{E_k + E_{k'} + \omega} 
                                               \ .  \label{eq:30}
\end{equation}
This result is identical to that of Phillips and Wallace~\cite{PW96}.
We observe here that the elimination of the negative energy components of the
`equal-time' free Green's function, i.e. $\langle\,G_0\,\rangle \rightarrow
\langle\,G_0^{++}\,\rangle$  reduces the above expression for the `equal-time'
Green's function to terms with $\langle\,G_0^{++}\,\rangle$ only, {\it i.e.}
\begin{equation}
\langle\,G_0\,I^{(2)}\,G_0\,\rangle = 2\langle\,G_0^{++}\,\rangle\,d_+\,
       \langle\,G_0^{++}\,\rangle\ .                 \label{eq:31}
\end{equation}
As a result, the potential in the Klein equation reduces to the simple form
used in time ordered perturbation theory:
\begin{equation}
K^{(2)}_1({\bf k},{\bf k'};s) \approx \frac{g^2}{\omega}\,
       \frac{1}{\sqrt{s}-E_k-E_{k'}-\omega+i\epsilon}
       \ .                                          \label{eq:32}
\end{equation}
We will refer to this approximation as the Klein potential with no 
anti-particles (Klein-NAP).
This major simplification of the potential, in neglecting the negative
energy component of the two-particle free Green's function, can have 
a major reduction in the structure of the crossed two $\sigma$
exchange contribution. 

Before we proceed to the evaluation of potential $K_1$ to order $g^4$, 
we need to estimate the contribution of the negative energy component of the 
$\phi$--$\phi$ Green's function. 
To get our scalar model to be a reasonable approximation
to the nucleon--nucleon interaction with pion exchange, we have chosen the 
mass of the $\phi$ to be $m=1.0$~GeV, while the mass of the $\sigma$ is taken  
as $\mu=0.15$~GeV. In this way the range of the interaction is comparable to 
that generated by one pion exchange.
In Fig.~\ref{fig.1} we plot the mass of the $\phi$--$\phi$ bound state 
as a function of the strength of the coupling, {\it i.e.} 
$\frac{g^2}{4\pi}$, for the BS, Klein, and Klein-NAP. 
Here we observe that the negative energy states can 
make a substantial contribution to the binding energy if the coupling 
is strong enough. However, if we assume that the binding energy of the 
$\phi$--$\phi$ system is comparable to that of the deuteron 
({\it i.e.} 2.225~MeV), then the coupling constant is 
$\frac{g^2}{4\pi}=1.646$, and the contribution to the binding energy 
of the negative energy states or anti-particles is negligible. In 
Fig.~\ref{fig.2} we present the phase shifts for the one sigma
exchange potential, corresponding to the coupling constant of 
$\frac{g^2}{4\pi}=1.646$, for the BS, Klein, 
and Klein-NAP equations as a function of energy. Here we note 
that all three results are very close, and that the contribution of the 
negative energy states is comparable to the difference between the BS 
and  the Klein equations. The fact that the Klein equation gives a 
good approximation to the BS equation is an indication that one can work 
within the framework of the three-dimensional Klein equation with no
anti-particles for the range of coupling constants that are consistent
with the nucleon-nucleon potential.

At this stage we have carried out the determination of the potential for
the Klein equation to order $g^2$. To include the crossed $\sigma$-exchange,
we need to go to fourth order in the coupling constant. The starting point 
is still Eq.~(\ref{eq:16}), but now we have to include all contributions
to $K$ to order $g^4$ as well as the contribution from $G_{K_2}$. With 
the help of Eq.~(\ref{eq:14}), we can write $K_1$, after one iteration of
$G_{K_2}$, as:
\begin{eqnarray}
K_1 &=& \langle\,G_0\,\rangle^{-1}\ \langle\,G_0\,K\,G_{K_2}\,\rangle\ 
         \langle\,G_0\,\rangle^{-1}                   \nonumber\\
    &\approx& \langle\,G_0\,\rangle^{-1}\ \langle\, G_0\,K\,\left[\,G_0 + 
              G_0\,(K-K_1)\,G_0\,\right]\,\rangle\ 
             \langle\,G_0\,\rangle^{-1}\ .        \label{eq:33}  
\end{eqnarray}
If we now take the kernel of the BS equation to fourth order in the 
coupling, {\it i.e.}
\begin{equation}
K \approx I^{(2)} + I^{(4)}\ ,                    \label{eq:34}
\end{equation}
where $I^{(4)}$ is the crossed $\sigma$-exchange diagram, and keep terms 
to fourth order in the coupling, we get
\begin{eqnarray}
K_1 &\approx& \langle\,G_0\,\rangle^{-1}\ \langle\,G_0\,I^{(2)}\,G_0\,\rangle\ 
            \langle\,G_0\,\rangle^{-1} 
           + \langle\,G_0\,\rangle^{-1}\ \langle\,G_0\,I^{(4)}_x\,G_0\,\rangle\ 
            \langle\,G_0\,\rangle^{-1} \nonumber \\
    &&\quad + \langle\,G_0\,\rangle^{-1}\,
            \left[\,\langle\,G_0\,I^{(2)}\,G_0\,I^{(2)}\,
           G_0\,\rangle - \langle\,G_0\,I^{(2)}\,G_0\,K_1^{(2)}\,G_0\,\rangle\,
           \right]\,\langle\,G_0\,\rangle^{-1}\nonumber\\
   &\equiv& K_1^{(2)} + K_1^{(4)}        \ .      \label{eq:35}
\end{eqnarray}
The first term is just the `equal-time' single $\sigma$-exchange, while the
second term is the `equal-time' crossed $\sigma$-exchange. In addition to
these two contributions, we have a part of the boxed diagram that cannot
be represented by the sequential `equal-time' two $\sigma$-exchange. This
contribution is given in the third and final term, and consists of the 
`equal-time' boxed diagram minus the once iterated `equal-time' single
$\sigma$-exchange, since
\begin{equation}
\langle\,G_0\,\rangle^{-1}\ \langle\,G_0\,I^{(2)}\,G_0\,K_1^{(2)}\,G_0\,
       \rangle\  \langle\,G_0\,\rangle^{-1} 
     = K^{(2)}_{1}\ \langle\,G_0\,\rangle\ K^{(2)}_{1}\ . \label{eq:36}
\end{equation}
In general, the evaluation of this potential, $K_1^{(4)}$, involves the
evaluation of the relative energy integration in both initial and final states
for both the boxed and crossed $\sigma$-exchanges. To facilitate the 
evaluation of these integrals, we have retained only the positive energy
component of the free $\phi$--$\phi$ Green's function, {\it i.e.} we have 
assumed that $G_0\rightarrow G_0^{++}$ in the evaluation of the potential. 
This gives us the result of time ordered perturbation theory~\cite{R91,L95},
{\it i.e.} the potential $K_1^{(4)}$ is the sum of the diagrams in 
Fig.~\ref{fig.3}. Note that in making use of time ordered perturbation theory
we have assumed that the contribution of anti-particles is negligible. 
That in turn was justified on the basis of our results for the present 
Lagrangian and at a value of the coupling constant that gives a binding 
energy for the $\phi$--$\phi$ that is comparable to the deuteron binding 
energy.

\section{Numerical Results}\label{sec.4}

To examine the approximation required to reduce the BS equation to the
non-relativistic Schr\"odinger equation with a local coordinate space potential,
we define the interaction at the level of the BS equation. We will include
in the interaction either a single $\sigma$-exchange ({\it i.e.} order $g^2$) or 
$\sigma$-exchange plus crossed $\sigma$-exchange ({\it i.e.} order $g^4$).
At this stage the only parameter in the potential is the coupling constant 
$g$, which we
set in the last section to reproduce a $\phi-\phi$ bound state with a 
binding energy of 2.225~MeV. This fixes the coupling constant at a value of
$\frac{g^2}{4\pi}=1.646$ for single $\sigma$-exchange, and 
$\frac{g^2}{4\pi}=1.484$ for single $\sigma$-exchange and crossed $\sigma$
exchange. We now carry through a number of approximations to
this fully covariant model. These being: (i)~At the one $\sigma$-exchange
level, we compare the results of the BS with the Klein-NAP, and the 
corresponding potential with no energy dependence. 
(ii)~At the $\sigma$-exchange plus crossed $\sigma$-exchange, we compare the
results of the BS and Klein-NAP with the results of removing first the 
energy dependence, {\it i.e.} the retardation, and second the
momentum dependence, {\it i.e.} the non-locality of the potential in the 
Klein equation.

Since the Klein-NAP approximation is identical to time ordered perturbation
theory, the potential has energy dependence of the form exhibited in 
Eq.~(\ref{eq:32}). This energy dependence corresponds to a retardation
in the potential which may cause problems if implemented in many-body
calculations. To remove the energy dependence in a typical denominator 
{\it e.g.}
\begin{equation}
D = \frac{1}{\sqrt{s} - E_k - E_{k'} - \omega}    \ , \label{eq:37}
\end{equation}
we first expand the energy of the $\phi$ and the total energy of the 
system in a power series of the inverse of the $\phi$ mass on the grounds that 
the typical momenta are small when compared with the $\phi$ mass, {\it i.e.}
\begin{eqnarray}
\sqrt{s} &=& 2\,\sqrt{k_0^2 + m^2} = 2m + \frac{k_0^2}{m} + \cdots \nonumber \\
E_k      &=& \sqrt{k^2 + m^2} = m + \frac{k^2}{2m} + \cdots
             \ ,                                      \label{eq:38}
\end{eqnarray}
where $k_0$ is the on-shell momentum. We then follow Sugawara and 
Okubo~\cite{SO60} and Rijken~\cite{R91} and expand $D$ in powers of 
$m^{-1}$, with the result that
\begin{equation}
D = -\frac{1}{\omega}\,\left[\,1 + \frac{2k_0^2-k^2 -k'^2}{2m\omega}
   + \left(\frac{2k_0 - k^2 - k'^2}{2m\omega}\right)^2 
   + \cdots\,\right]\ .                               \label{eq:39}
\end{equation}
The first term in this expansion, {\it i.e.} $O(1)$, is referred to as the
``adiabatic'' approximation, and if we keep terms of $O(m^{-1})$ we have the
``non-adiabatic'' approximation~\cite{R91}. Although this 
approximation might be valid for the $\sqrt{s}$ below the threshold for 
$\sigma$ production, the corresponding expansion for $E_{k'}$ can be 
questioned since $k'$ is an integration variable that takes on all 
values from zero to infinity in solving the integral 
equation~\cite{BJK69}. The ``adiabatic'' approximation at the one 
$\sigma$-exchange level is basically 
the static approximation in which $E_k + E_{k'} = \sqrt{s}$, and in this 
limit the potential has no energy dependence, and  reduces to a local 
Yukawa potential. 

For the two $\sigma$-exchange contribution, {\it i.e.} order $g^4$, 
we get factors of the form given
in Eq.~(\ref{eq:37}) for intermediate states. In this case to remove the 
energy dependence one needs to carry through an ``on-energy-shell'' 
approximation which requires that:
\begin{equation}
k^2 - k_0^2 \rightarrow 0\quad\mbox{and}
\quad k'^2-k_0^2\rightarrow 0\ ,                    \label{eq:40}
\end{equation}
where $k$ and $k'$ are the initial and final state momenta respectively. 
The final form of the resultant potential\cite{R91,L95} to order $g^4$ in 
both the adiabatic and non-adiabatic approximations, has no energy dependence. 
The momentum dependence then is determined by the expansion in powers of
$m^{-1}$ of all energy denominators, and the number of terms kept in such an
expansion.

In Fig.~\ref{fig.4} we present the $s$-wave phase shifts for the case when the
potential is defined to be due to a single $\sigma$-exchange. Here we 
observe that although the results of the BS and Klein-NAP equations 
are in reasonably good agreement, the removal of the energy dependence has a 
substantial effect on the phase shifts. Although this difference could be 
compensated for by the adjustment of the coupling constant $g$, it is not 
clear that the off-energy-shell difference will not persist. To illustrate
this, we present in Fig.~\ref{fig.5} the Kowalski-Noyes\cite{K65,N65}
half-off-shell function for the Klein-NAP potential (solid line) and the energy 
independent potential I (dashed line). Also included, is the result for the 
half-off-shell function for an energy independent potential II 
(dotted line) that gives almost the same phase shifts as the Klein-NAP 
potential. Clearly the half-off-shell function for the energy independent 
and energy dependent potentials are quite different. This is true even 
when the two potentials have approximately the same phase shifts. 
Such a large variation in the off-shell behavior of the amplitude, 
even when the potentials are identical on-shell, can have significant effects 
on three- and many-body results~\cite{MSS96}.

To get a measure of the uncertainty in the coupling constant $g$ as a 
result of removing the energy dependence of the potential while
keeping the physical observables the same, we have
adjusted the coupling constant $g$ such that the binding energy of
the $\phi$--$\phi$ system is 2.225~MeV in the BS, Klein-NAP and the
energy independent potential. The resultant coupling constants are 
presented in Table~\ref{table.1}. Here we observe a change of 6\% in
going from the BS to the Klein-NAP equation as compared with a 13\% 
change in going from the BS to energy independent potential.
This change in the coupling constant is significant if the coupling 
constant is to be a meaningful quantity that might be derived from some 
underlying quantum field theory, such as QCD for the nucleon-nucleon 
interaction.

The contribution to order $g^4$ for the potential in the Klein-NAP 
approximation has both non-locality in the form of momentum dependence,
and a dependence on the energy. By taking the initial and final momenta
on the energy shell, we remove the energy dependence, but not the 
non-locality. This non-locality is removed by implementing the procedure
in Eq.~(\ref{eq:38}). The resultant potential to order $g^4$ is now
local~\cite{R91,RS96}, and can be used in the three- and many-body
calculations. In Fig.~\ref{fig.6} we present the phase shifts for potentials
calculated to order $g^4$. Included are the phase shifts for the BS, 
Klein-NAP, and the energy independent potentials to $O(1)$ (adiabatic) 
and $O(m^{-1})$ (non-adiabatic). Here again, the results of the BS and
Klein-NAP are in reasonable agreement, but the energy independent 
approximations are substantially different. The fact that the results 
for the adiabatic and non-adiabatic approximations are different is an 
indication that the series expansion in $m^{-1}$ is not as convergent 
as we would like it to be. Here again
the off-shell behavior of the amplitude for the Klein-NAP and energy
independent potentials are substantially different as illustrated in 
Fig.~\ref{fig.7} where the Kowalski-Noyes~\cite{K65,N65} function is given
for these potentials. As was the case with the phase shifts, the non-adiabatic
approximation is substantially better than the adiabatic approximation.

In the nucleon-nucleon interaction, the $\pi NN$ vertex has associated with
it a form factor that is a function of the exchange pion momentum. This 
form factor is introduced to overcome the singular nature of the potential.
For the Lagrangian under consideration the potential is not singular and
there is no need for any form factors. However, to examine the effect of such
a form factor on the difference between the energy dependent and energy 
independent potentials, we have introduced a Gaussian form factor\cite{R91} 
by the substitution
\begin{equation}
g^2 \rightarrow g^2\,e^{-k^2/\Lambda^2}\ ,          \label{eq:41}
\end{equation}
where $k$ is the momentum of the exchanged $\sigma$ meson. In 
Fig~\ref{fig.8} we illustrate the changes in the phase shifts for the 
Klein-NAP and energy independent $\sigma$-exchange potentials as a function 
of the cut-off parameter $\Lambda$ at a laboratory energy of 100~MeV. 
Here we observe that as the cut-off mass is reduced, the difference between
energy dependent and energy independent solutions is also partly reduced. 
This suggests that the energy dependence is a short range
effect, and as the form factor starts to dominate the short range behavior
of the potential, the role of the energy dependence is suppressed.

\section{Conclusion}\label{sec.5}

In an attempt to understand the approximation involved in reducing the
potential in the Bethe-Salpeter (BS) equation to an equivalent local 
potential for use in the Schr\"odinger equation, we have considered a 
series of approximations which involved: 
(i)~The reduction of the equation from four- to three-dimensions.
(ii)~The elimination of negative energy states or anti-particles.
(iii)~The removal of energy dependence and non-locality in the potential.
To maintain  simplicity in the model, while maintaining some relevance  
to the nucleon-nucleon interaction, we have considered a 
scalar $\phi^2\sigma$ interaction Lagrangian and included in the 
potential all diagrams to order $g^4$, {\it i.e.} single $\sigma$-exchange
and crossed two $\sigma$-exchange in analogy with one- and two-pion 
exchanges in the nucleon-nucleon problem. To set the strength of this interaction,
we required that the $\phi$--$\phi$ system have a bound state energy 
comparable to the deuteron binding energy.

In reducing the two-body problem from four- to three-dimensions, we have 
taken advantage of the equal-time Green's function~\cite{LT63} and the
requirement that the potential have no dependence on the relative 
energy~\cite{K53}, to derive a scheme that allows for a systematic way of
improving the potential to reproduce the results of the BS equation~\cite{PW96}.
We then found that at the one $\sigma$-exchange level, neglecting the 
contribution of negative energy states is not substantial, provided that
the $\phi$--$\phi$ system has a weakly bound state as in the deuteron. 
The elimination of negative energy states gives us the time-ordered 
perturbation theory that has been used as the starting point for the recent
derivation of the nucleon-nucleon potential~\cite{MHE87,R91,RS96}.
At this level there is good agreement between the BS equation and 
the corresponding three-dimensional equation we refer to as the
Klein-NAP~\cite{K53}. However, the potential now is both energy dependent,
{\it i.e} has retardation, and is non-local. These features give rise
to problems if the potential is to be used for three- and many-body 
calculations. To avoid such problems, it has become standard practice to
remove the energy dependence and non-locality. Here we find that the
approximations required to remove this energy dependence and non-locality 
results in a drastic change in the phase shifts.

If we adopt the view that the coupling constant can, at the local potential
level, be adjusted to fit the experimental phase shifts, then the
coupling constant defined at the covariant level could change by as 
much as 10-15\%. This raises a problems in the nucleon-nucleon case if
the $\pi NN$ coupling constant extracted from the experimental 
data is to be compared with theoretical results extracted from QCD.
This problem becomes central when the fit to the experimental phase 
shifts is at a level where the $\chi^2$ per data is near one.
More interesting is the fact that the
off-shell behavior of the final local potential is substantially 
different from the original potential derived from the time-ordered 
perturbation theory. This change in off-shell behavior has significant 
ramifications in the three- and many-body results based on local
potentials, and can be the origin of the three-nucleon force introduced
to get the correct binding energy of light nuclei.

Although we have considered one of an infinite set of 3-D equations 
with the kernel calculated to order $g^{4}$, we should point out that 
for the bound state problem with the value of the coupling constant 
used here, the results of the Klein equation are in good agreement with 
the solution of the BS equation in which the kernel is calculated to 
all orders~\cite{PW96,NT95,N95}. Finally, we should point out that there 
have been similar analyses based on the Blankenbecler-Sugar 
equation~\cite{FJ71,MG78}.

\acknowledgments

The authors would like to thank Daniel Phillips and Vincent Stoks for their
interest and help over the duration of this investigation, Coen 
van~Antwerpen and Daniel Phillips for the Bethe-Salpeter code, and Daniel 
Phillips for his comment on the initial draft of this manuscript. Finally 
we also would like to thank the Australian Research Council for their 
financial support without which the project would not have been possible.


\newpage

\begin{figure}
\caption{The bound state energy of the $\phi-\phi$ system in the Bethe-Salpeter, 
Klein, and Klein with no anti-particle as a function of the coupling strength
$\frac{g^2}{4\pi}$}\label{fig.1}
\end{figure}


\begin{figure}
\caption{The phase shifts for the $\phi-\phi$ system in the Bethe-Salpeter, 
Klein, and Klein with no anti-particles as a function of the energy for
$\frac{g^2}{4\pi}=1.646$. This corresponds to a binding energy comparable
to the deuteron binding energy.}\label{fig.2}
\end{figure}


\begin{figure}
\caption{The diagrams that contribute to the potential $K_1^{(4)}$  to fourth
order in  the coupling constant in time ordered perturbation theory. Diagrams
(a) and (b) are the single $\sigma$-exchange, {\it i.e.} $K_1^{(2)}$, while
diagrams (e) -- (j) are the crossed diagrams. Here diagrams (c) and (d) are
those diagrams that are in the iterated $\sigma$-exchange in the BS equations
but are not generated in the iteration of the Klein equation and are part of
the fourth order contribution to the potential.}\label{fig.3}
\end{figure}


\begin{figure}
\caption{The $s$-wave phase shifts for the BS, Klein-NAP, and energy independent
$\sigma$-exchange potentials.}\label{fig.4}
\end{figure}


\begin{figure}
\caption{The Kowalski-Noyes half-off-shell function for the Klein-NAP and
the energy independent $\sigma$-exchange potentials. The energy independent I
corresponds to the potential that gave the phase shifts in Fig.~\ref{fig.4},
while energy independent II corresponds to the case when the coupling constant
$g$ is adjusted to give the same phase shifts for the energy independent and
the Klein-NAP potentials.}\label{fig.5}
\end{figure}


\begin{figure}
\caption{The $s$-wave phase shifts for the BS, Klein-NAP,  adiabatic 
$O(1)$, non-adiabatic $O(m^{-1})$ single
$\sigma$-exchange and crossed sigma potentials.}\label{fig.6}
\end{figure}


\begin{figure}
\caption{The Kowalski-Noyes half-off-shell function for the Klein-NAP,  
adiabatic $O(1)$, non-adiabatic $O(m^{-1})$ single
$\sigma$-exchange and crossed sigma potentials.}\label{fig.7}
\end{figure}


\begin{figure}
\caption{The $s$-wave phase shifts at $E_{lab}=100$~MeV for the
Klein-NAP and energy independent potential for the $\sigma$-exchange 
potential.}\label{fig.8}

\end{figure}

\newpage


\begin{table}
\caption{Comparison of the coupling constant required to give a binding
energy of 2.225~MeV for the $\phi$--$\phi$ system.\label{table.1}} 

\begin{tabular}{lc}  
Potential          & $\frac{g^2}{4\pi}$ \\ \hline
Bethe-Salpeter     & 1.646              \\
Klein-NAP          & 1.747              \\
Energy Independent & 1.433              \\ 
\end{tabular}
\end{table}

\newpage


\begin{references}
\bibitem{ORK96} C. Ord\'o\~nez, L. Ray and U. van~Kolck, Phys. Rev.
C {\bf 53}, 2086 (1996).
\bibitem{MHE87} R.~Machleidt, K.~Holinde, and Ch.~Elster, Phys. Rep. 
{\bf 149}, 1 (1987); R.~Machleidt, in {\it Advances in Nuclear Physics},
edited by J.~W.~Negele and E.~Vogt (Plenum, New York, 1989), Vol. 19,
pp. 189-376.
\bibitem{SKT94} V. G. J. Stoks, R. A. M. Klomp, C. P. F. Terheggen, and
J.~J.~de~Swart, Phys. Rev. C {\bf 49}, 2950 (1994).
\bibitem{SB51} E.~Salpeter and H.~Bethe, Phys. Rev. {\bf 84}, 1232 (1951).
\bibitem{LS50} B.~Lippmann and J. Schwinger, Phys. Rev. {\bf 79}, 
469 (1950).
\bibitem{KL74} A.~Klein, and T.-S.~H.~Lee, Phys. Rev. D {\bf 10}, 4308
(1974).
\bibitem{BbS66} R.~Blankenbecler, and R.~Sugar, Phys. Rev. {\bf 142},
1051 (1966).
\bibitem{G82} F.~Gross, Phys. Rev. C {\bf 26}, 2203 (1982).
\bibitem{K53} A.~Klein, Phys. Rev. {\bf 90}, 1101 (1953).
\bibitem{KP91} B.D.~Keister and W.N.~Polyzou, in {\it Advances in Nuclear
Physics}, Eds. J.W.~Negele and E.W.~Vogt (Plenum, New York), {\bf 20}, 
p.225 (1991).
\bibitem{Ok54} S. Okubo, Prog. Theor. Phys. {\bf 20}, 603 (1954).
\bibitem{Fu92} M.G.~Fuda, Nucl. Phys. {\bf A543}, 111c (1992);
               Ann. Phys. (N.Y.) {\bf 231}, 1 (1994);
M.G.~Fuda and Y.~Zhang, in {\it Proceeding of the XIVth International
Conference on Few Body Problem in Physics}, Ed. Franz Gross, AIP Conf. Proc.
No. 334 (AIP, New York), 879 (1995);
               Phys. Rev. C {\bf 51}, 23 (1995);
M.G.~Fuda, Phys. Rev. C {\bf 52}, 1260 (1995);
           Phys. Rev. C {\bf 54}, 495 (1996).
\bibitem{LT63} A.~A.~Logunov, and A.~N.~Tavkhelidze, Nuovo Cimento {\bf 29}
380 (1963).
\bibitem{PW96} D.~R.~Phillips and S.~J.~Wallace, Phys. Rev. C {\bf 54}, 
507 (1996).
\bibitem{IZ85} C. Itzykson and J. Zuber, {\it Quantum Field Theory}, 
(McGraw-Hill, Singapore, 1985).
\bibitem{R91} Th.A.~Rijken, Ann. Phys. (N.Y.) {\bf 208}, 253 (1991).
\bibitem{RS96} Th.A.~Rijken and V.G.J.~Stoks, Phys. Rev. C {\bf 46},
73 (1992); {\it ibid} C {\bf 46}, 102 (1992); {\it ibid} C {\bf 54},
2851 (1996); {\it ibid} C{\bf 54}, 2869 (1996).
\bibitem{KB94} A.~N.~Kvinikhidze and B. Blankleider, Few-Body Syst. Suppl. 
{\bf 7}, 294 (1994).
\bibitem{P95} D.~R.~Phillips, Ph.D. thesis, Flinders University of 
South Australia, 1995.
\bibitem{SR97} V.G.J. Stoks and Th.A. Rijken, Nucl. Phys. {\bf A613},
311 (1997).
\bibitem{L95} A.D. Lahiff, Honours Thesis, Flinders University of
South Australia (1995).
\bibitem{SO60} M. Sugawara and S. Okubo, Phys. Rev. {\bf 117},
605 (1960).
\bibitem{BJK69} G.E. Brown, A.D. Jackson, and T.T.S. Kuo, Nucl. Phys. 
{\bf A133}, 481 (1969).
\bibitem{K65} K. Kowalski, Phys. Rev. Lett. {\bf 15}, 798 (1965).
\bibitem{N65} H.P. Noyes, Phys. Rev. Lett. {\bf 15}, 538 (1965).
\bibitem{MSS96} R. Machleidt, F. Sammarruca, and Y. Song, Phys. Rev. C 
{\bf 53}, 1483 (1996).
\bibitem{NT95} T. Nieuwenhuis, and J.A. Tjon, Phys. Lett. {\bf B355},
               283 (1995); Phys. Rev. Lett. {\bf 77}, 814 (1996).
\bibitem{N95} T. Nieuwenhuis, Ph.D. Thesis, University of Utrecht (1995).

\bibitem{FJ71} M. Fortes and A.D. Jackson, Nucl. Phys. {\bf A175},
               449 (1971).
\bibitem{MG78} L. M\"uller and W. Glockle, Nucl. Phys. {\bf B146},
               393 (1978).
\end{references}
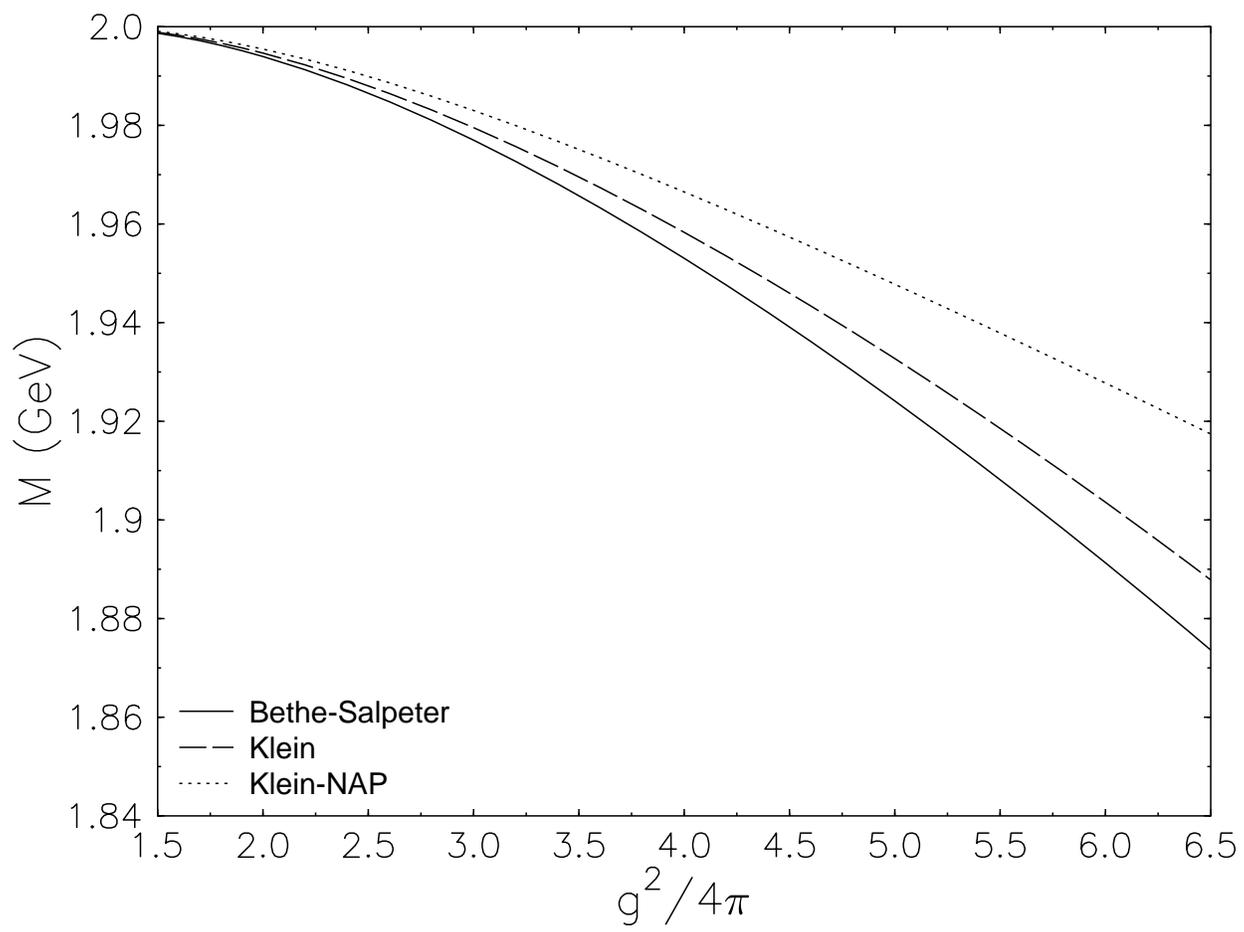
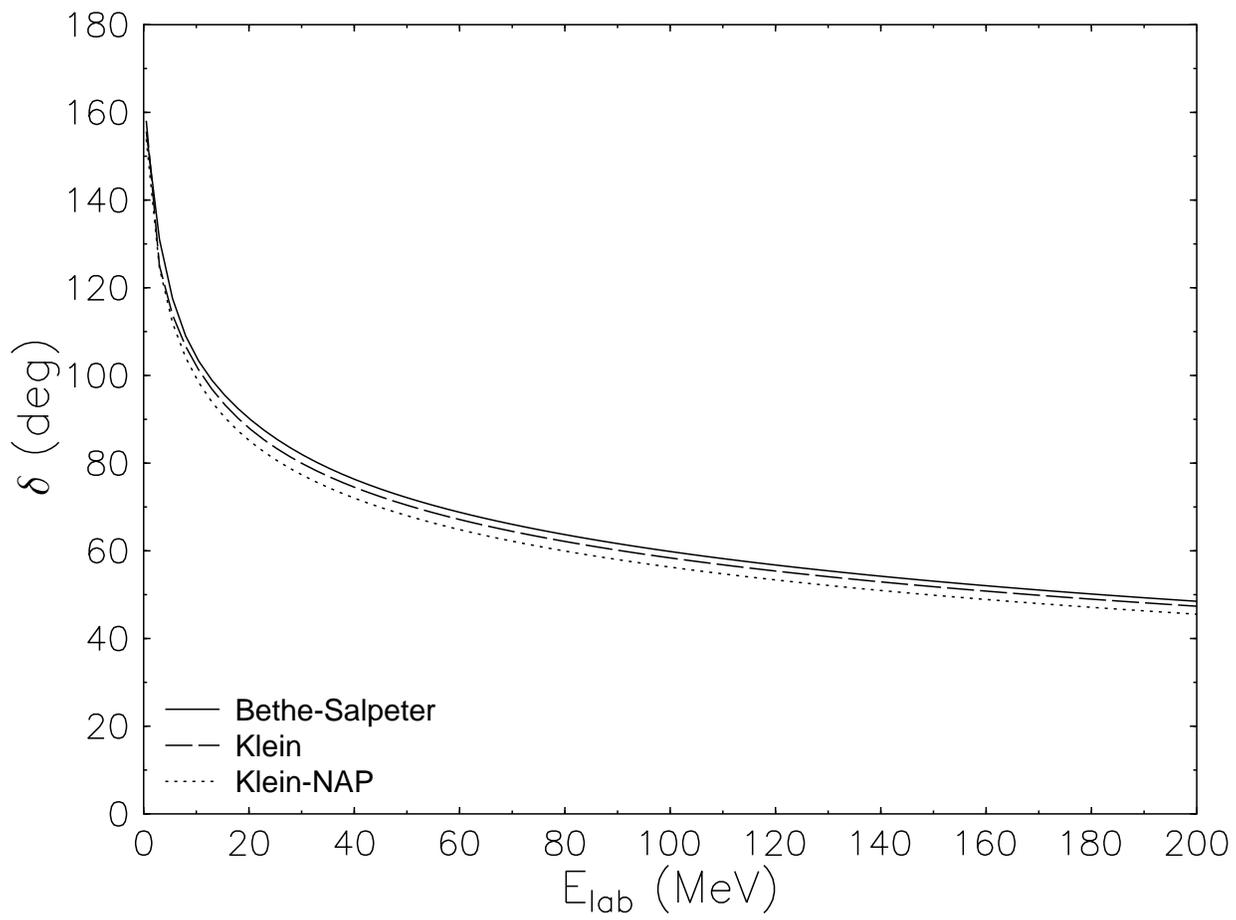
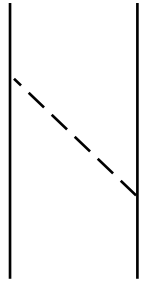
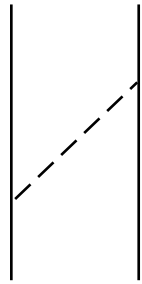
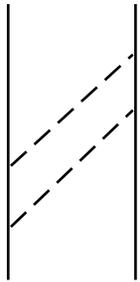
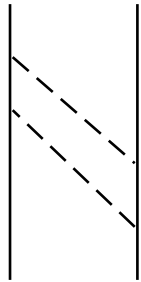
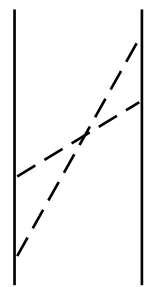
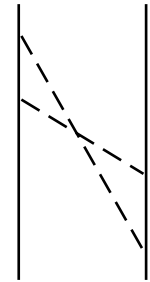
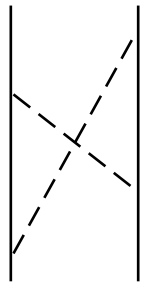
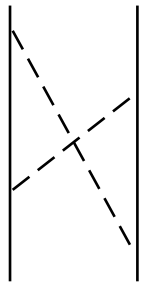
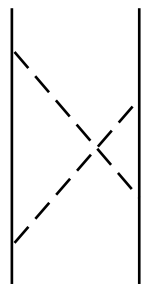
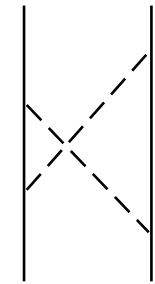
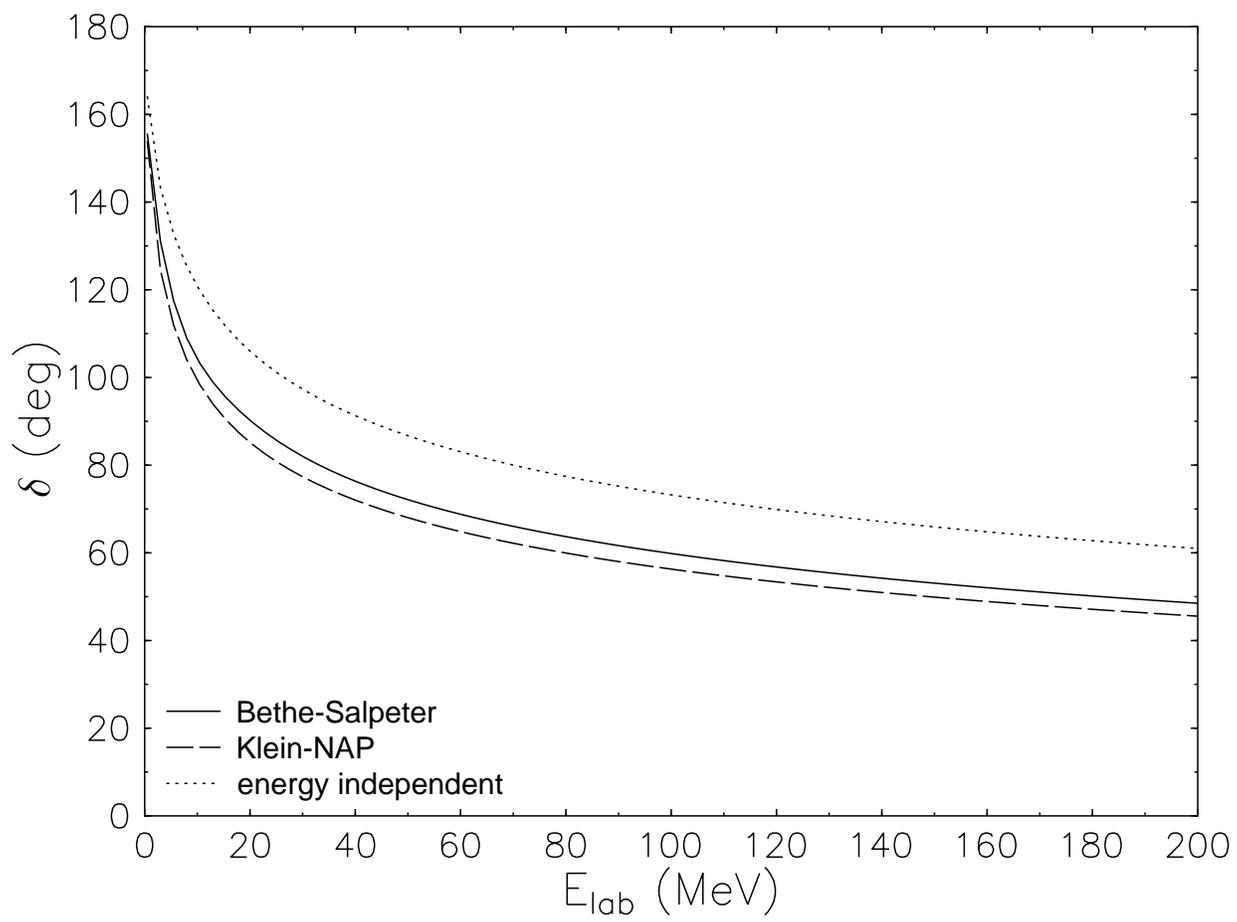
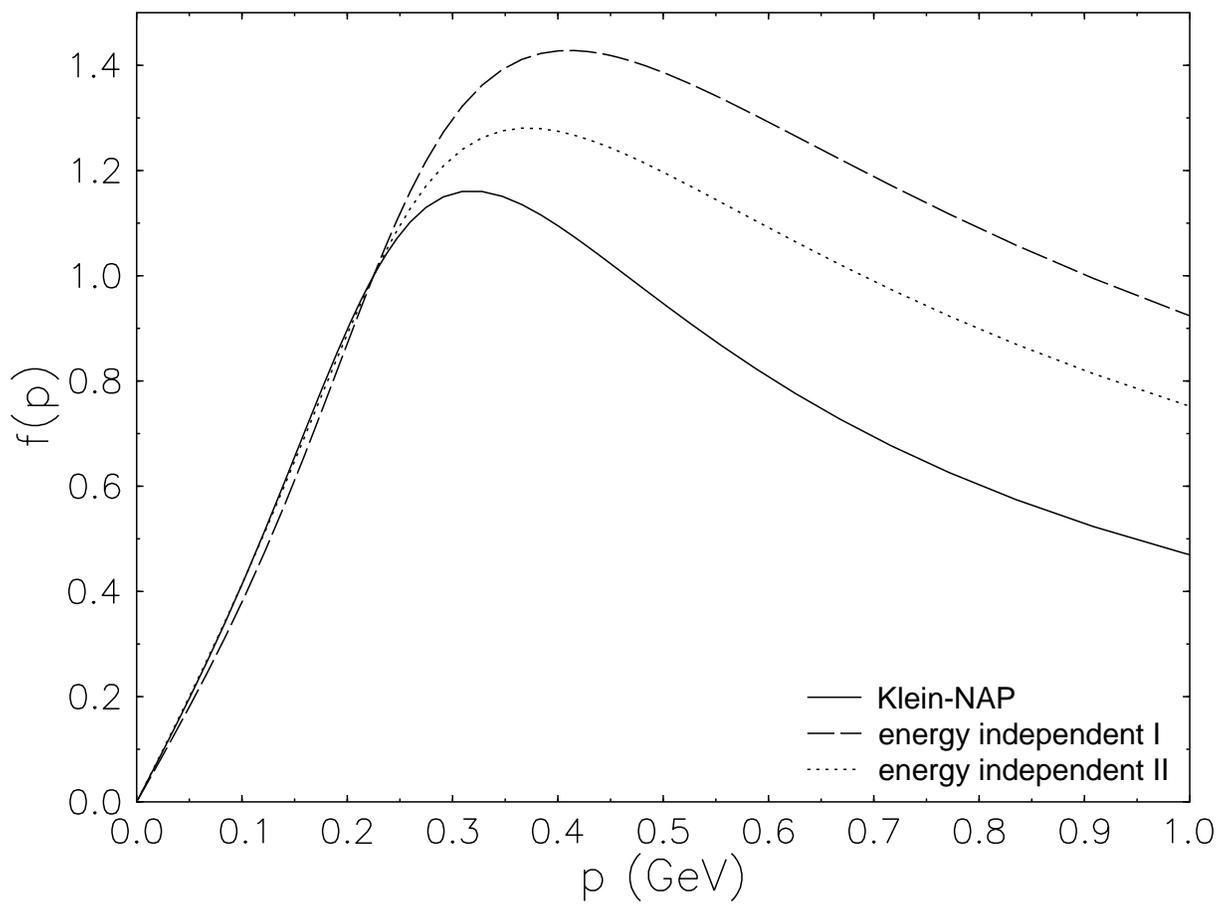
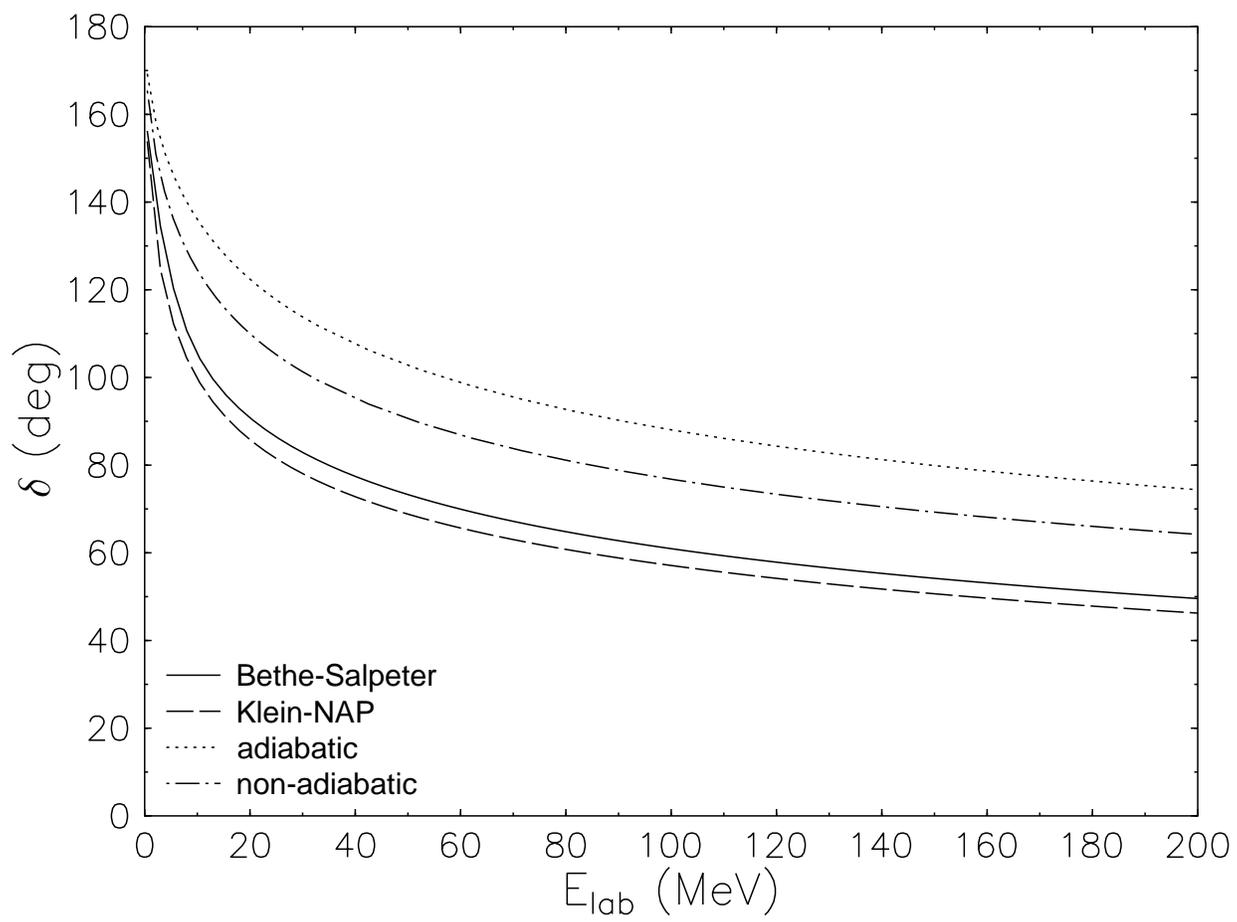
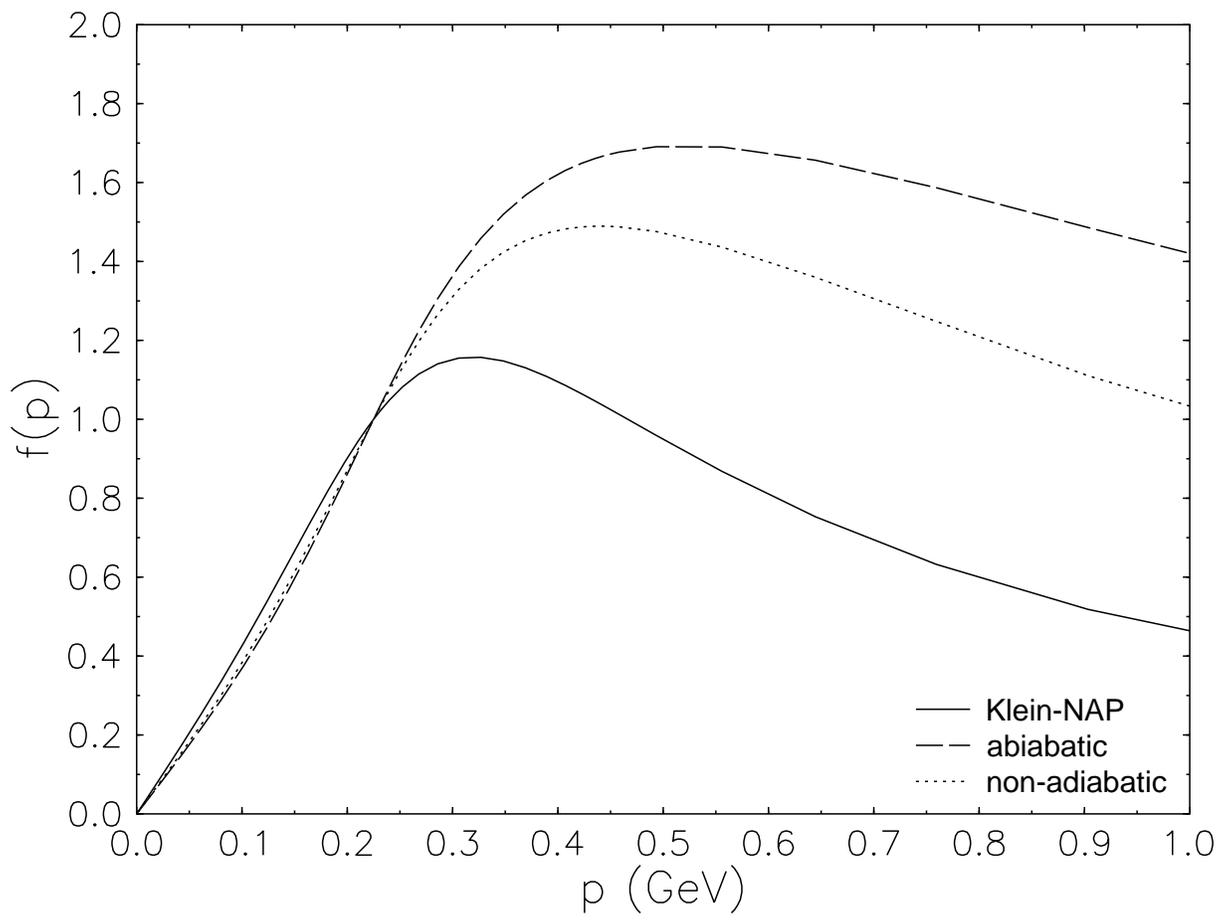
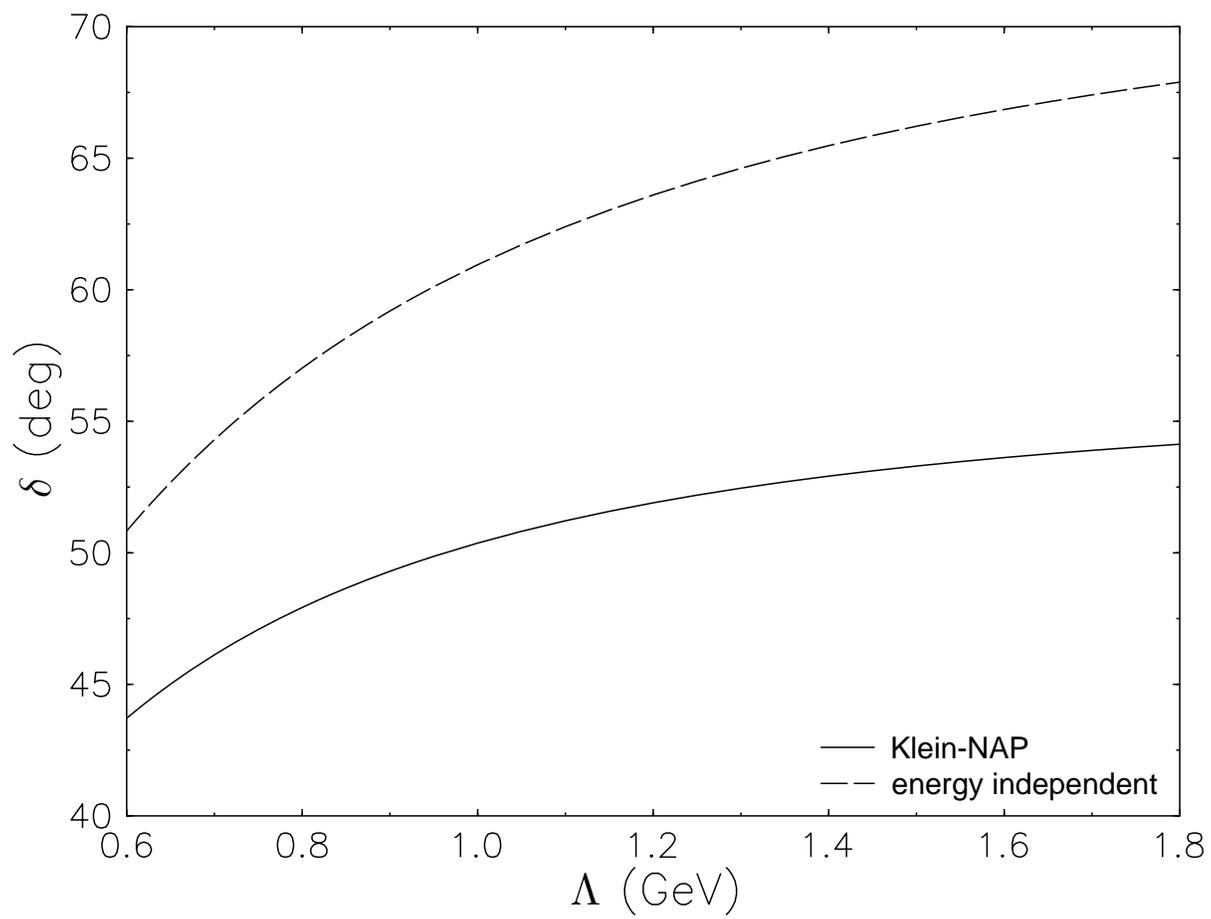
\end{document}